\documentclass{rspublic}
\usepackage{amssymb,amsmath}

\begin{document}

\title[Integrability and Linearization]{On the complete integrability and 
linearization of nonlinear
ordinary differential equations - Part II: Third order equations}

\author[Chandrasekar, Senthilvelan and Lakshmanan]{V. K. Chandrasekar, 
M. Senthilvelan and M. Lakshmanan}

\affiliation{Centre for Nonlinear Dynamics, Department of Physics, 
Bharathidasan Univeristy, Tiruchirapalli - 620 024, India}

\label{firstpage}

\maketitle

\begin{abstract}{Integrability, Integrating factor, Linearization, Equivalence
problem}
We introduce a method for finding general solutions of third-order nonlinear
differential equations by extending the modified Prelle-Singer method. We
describe a procedure to deduce all the integrals of motion associated with the
given equation so that the general solution follows straightforwardly from these
integrals. The method is illustrated with several examples. Further, we propose
a powerful method of identifying linearizing transformations. The proposed
method not only unifies all the known linearizing transformations systematically
but also introduces a new and generalized linearizing transformation (GLT). In
addition to the above, we provide an algorithm to invert the nonlocal
linearizing transformation. Through this procedure the general solution for the
original nonlinear equation can be obtained from the solution of the linear 
ordinary differential equation.
\end{abstract}
                                                               
\section{Introduction}
In a previous paper (Chandrasekar \textit{et al.} 2005) we have discussed the 
complete integrability aspects of a class of
second-order nonlinear ordinary differential equations (ODEs) through a nontrivial
extension of the so called Prelle-Singer (PS) (Prelle \& Singer 1983; 
Duarte \textit{et al.} 2001) procedure. We have illustrated the
procedure with several physically interesting nonlinear oscillator examples. We
have also developed a straightforward algorithmic way to transform the given 
second order nonlinear ODE to a linear free particle equation, if it is 
linearizable. 

One of the questions raised at the final stage of our earlier work 
(Chandrasekar \textit{et al.} 2005) was that what are the
implications of the novel features which we introduced in the extended
Prelle-Singer procedure to obtain the second constant of motion (in the
case of second order ODEs) to third and higher order ODEs. To have
a closer look at the problem let us recall our earlier work briefly here. We
considered a second order ODE of the form
$\frac{d^2x}{dt^2}=\frac{P(t,x,\dot{x})}{Q(t,x,\dot{x})},
\;{ P,Q}\in \mathbb{C}{[t,x,\dot{x}]}$, and explored two
pairs of independent functions, say, $R_i$ and $S_i$, $i=1,2,$ associated with 
the underlying
ODE. These functions are nothing but the integrating factors and null forms,
respectively. Once these
two pairs of functions are determined (by solving an overdetermined system
of first order PDEs) then each pair leads to an independent integral of motion,
which can then be used to find the general solution for the given equation. 
Thus instead of integrating the first
integral and obtaining the general solution which is conventionally followed in the
literature we implemented some novel ideas in
the PS method such that one can construct the general solution for the given
equation in a self contained way and, in fact, our procedure works for a class 
of problems.

In the case of third order ODEs, one should have three independent integrals 
of motion
in order to establish the complete integrability. To deduce these three integrals
one should have three pairs of independent functions ($R_i$, $U_i$ and $S_i$), 
$i=1,2,3$. 
When we extend the PS procedure to third order ODEs we find that the 
determining equations
for the integrating factors and null forms provide either one or two integrals
of motion only straightforwardly. Again the hidden form of the functions
($R_3,U_3,S_3$) should be
explored in order establish the complete integrability of the given equation 
within the frame work of PS procedure. In this paper we describe a 
procedure to capture the required set of functions. {\it With the completion of 
this task we formulate a simple, straightforward and powerful method to 
solve a wide class of third order ODEs of contemporary literature}. 

We stress at this point that the
application of PS procedure to third order ODEs is not a straightforward
extension of the second order case. In fact  one has to overcome many faceted
problems. The first and foremost one is how to solve the determining
equations in such a way that one could obtain three sets of independent 
functions, namely,
$(R_i,U_i,S_i),\;i=1,2,3,$ in a systematic way. In the present case we have six
equations for three unknown functions (in the case of second order equations we have three
equations for two unknowns). We overcome this problem by adopting suitable
methodologies,
the details which we present in \S3. Another obstacle one could face in higher
order ODEs, at least in some cases, is that one may be able to get only one 
integral of 
motion and in this situation how one would be able to generate the
remaining integrals of motion from the first integral is also tackled by us in 
this paper. 

Our main goal, besides the above, is to bring out a novel and straightforward
way to construct linearizing transformations for third order ODEs. The latter
can be used to transform the given third order nonlinear ODEs to a linear
equation. We note that unlike the second order
equations, the third order ODEs can be linearized through different kinds of
transformations, namely, invertible point transformation 
(Ibragimov \& Meleshko 2005), contact transformation 
(Bocharov \textit{et al.} 1993; Ibragimov \& Meleshko 2005)
and generalised Sundman transformation 
(Berkovich \& Orlova 2000; Euler \textit{et al.} 2003; Euler \& Euler 2004). 

In this paper we introduce a new kind of transformation which can be
effectively used to linearize a class of nonlinear third order ODEs. In fact,
one can linearize certain equations only through this
transformation alone and not by the known ones in the literature. We call this
transformation as generalized linearizing transformation (GLT). We note that 
generalised Sundman transformation is a special case of this transformation. 
In the generalised Sundman transformation the new independent variable is a
nonlocal one and so eventhough one is able to transform the given nonlinear
third order ODE to a linear one due to the nature of nonlocal independent variable
it is not easy to write down the general solution. In the case of generalized 
linearizing transformation both the new dependent and independent  
variables contain derivative terms also in addition to the
independent variable being nonlocal. Even for this general case, in this paper,
we succeed to present an efficient algorithm to deduce the general solution.

Another fundamental problem regarding linearization is how to deduce the 
linearizing transformations systematically. Generally, Lie symmetry analysis and
direct methods are often used to deduce the point and contact transformations 
(Steeb 1993; Olver 1995; Bocharov \textit{et al.} 1993; Bluman \& Anco 2002; 
Ibragimov \& Meleshko 2005). In this work we 
propose a simple and straightforward method to
deduce linearizing transformations and we derive them from the first integral.
Our method of deducing linearizing transformations has several salient 
features. Irrespective of the
form of the linearizing transformation (point/contact/generalised Sundman 
transformation), it can be derived from the first
integral itself. We also note that one can also linearize a third order ODE to the
second order free particle equation
through our method. An added advantage of our method is that suppose a given
equation is linearizable through one or more kinds of transformations then our
procedure provides all these transformations in a straightforward way and as far as
our knowledge goes {\it no such single method has been formulated in the 
literature}. 

The plan of the paper is follows. In the following section we extend the PS
procedure to third order ODEs and indicate new features in finding the three
independent integrals of motion. In \S3, we describe the methods of solving the
determining equations and how one can obtain compatible solutions from them. We
illustrate the procedure with several examples. In \S4, we propose a powerful
method of identifying linearizing transformations. This method not only brings
out all the known transformations systematically but also a new generalized
linearizing transformation (GLT) for the third order ODEs. We emphasize the 
validity of the method
with several illustrative examples arising in different areas of mathematics
and physics. We present our conclusions in \S5.

\section{Prelle-Singer method for third order ODEs}
\label{sec2}
\subsection{General Theory}
Let us consider a class of third order ODEs of the form 
\begin{eqnarray} 
\dddot{x}=\frac{P}{Q},  \quad 
{ P,Q}\in \mathbb{C}{[t,x,\dot{x},\ddot{x}]}, \;\;( ^{.}=\frac{d}{dt})
 \label{met1}
\end{eqnarray}
where over dot denotes differentiation with respect to time and $P$ and $Q$ 
are polynomials in $t$, $x$, $\dot{x}$ and $\ddot{x}$ with coefficients in the 
field of complex numbers, $\mathbb{C}$. Let us assume that the third order 
ODE (\ref{met1}) 
admits a first integral $I(t,x,\dot{x},\ddot{x})=C,$ with $C$ being constant on the 
solutions, so that the total differential of $I$ gives
\begin{eqnarray}  
dI={I_t}{dt}+{I_{x}}{dx}+{I_{\dot{x}}{d\dot{x}}}+{I_{\ddot{x}}{d\ddot{x}}}=0, 
\label{met3}  
\end{eqnarray}
where each subscript denotes partial differentiation with respect 
to that variable. Equation~(\ref{met1}) can be rewritten as 
$\frac{P}{Q}dt-d\ddot{x}=0$. Now adding the null terms 
$U(t,x,\dot{x},\ddot{x})\ddot{x}dt-U(t,x,\dot{x},\ddot{x})d\dot{x} $ and 
$S(t,x,\dot{x},\ddot{x})\dot{x}dt-S(t,x,\dot{x},\ddot{x})dx $
to this we obtain that, on the solutions, the 1-form
\begin{eqnarray}
(\frac{P}{Q}+S\dot{x}+U\ddot{x})dt-Sdx-Ud\dot{x}-d\ddot{x} = 0. 
\label{met6} 
\end{eqnarray}
	
Looking at equations~(\ref{met3}) and 
(\ref{met6}) one can conclude that, on the solutions, these two forms are 
proportional and the form of equation~(\ref{met6}) is equivalent to equation~(\ref{met3})
except for an overall multiplication factor. Thus multiplying equation~(\ref{met6}) by 
the factor $ R(t,x,\dot{x},\ddot{x})$ which acts as the integrating factor
for (\ref{met6}), we have on the solutions that 
\begin{eqnarray} 
dI=R(\phi+S\dot{x}+U\ddot{x})dt-RSdx-RUd\dot{x}-Rd\ddot{x} =0, 
\label{met7}
\end{eqnarray}
where $ \phi\equiv {P}/{Q}$. Comparing equations (\ref{met3}) 
with (\ref{met7}) we have the following relations, on the solutions, 
\begin{align}  
 I_{t}  = R(\phi+S\dot{x}+U\ddot{x}), \qquad
 I_{x}  = -RS, \qquad
 I_{\dot{x}}  = -RU,\qquad
 I_{\ddot{x}}  = -R.  
 \label{met8}
\end{align} 
Now imposing the compatibility conditions, 
$I_{tx}=I_{xt}$, $I_{t\dot{x}}=I_{{\dot{x}}t}$, $I_{t\ddot{x}}=I_{{\ddot{x}}t}$,
$I_{x{\dot{x}}}=I_{{\dot{x}}x}$, $I_{x{\ddot{x}}}=I_{{\ddot{x}}x}$,
$I_{\dot{x}{\ddot{x}}}=I_{{\ddot{x}}\dot{x}}$, which exist between the 
equations~(\ref{met8}), we have the following equations which constitute three
determining equations ((\ref{met9})-(\ref{met11}) given below) for the 
functions $S$, $U$ and $R$ along with three constraints 
((\ref{met12})-(\ref{met14}) given below) that they need to satisfy,
\begin{align}
D[S] & = -\phi_x+S\phi_{\ddot{x}}+US,\label{met9}\\
D[U] & = -\phi_{\dot{x}}+U\phi_{\ddot{x}}-S+U^2,\label{met10}\\
D[R] & = -R(U+\phi_{\ddot{x}}),\label{met11}\\
R_x & = R_{\ddot{x}}S+RS_{\ddot{x}},
\label{met12}\\
R_{\dot{x}}S & = -RS_{\dot{x}}+R_{x}U+RU_{x}, \label{met13}\\
R_{\dot{x}} & = R_{\ddot{x}}U+RU_{\ddot{x}}, \label{met14}
\end{align}
where 
\begin{eqnarray}
D=\frac{\partial}{\partial{t}}+
\dot{x}\frac{\partial}{\partial{x}}+\ddot{x}\frac{\partial}{\partial{\dot{x}}}
+\phi\frac{\partial}{\partial{\ddot{x}}}.\label{opp}
\end{eqnarray}
The task of solving equations~(\ref{met9})-(\ref{met14}) can be 
accomplished in the following way.  
Substituting the given expression of $\phi$ into (\ref{met9})-(\ref{met10}) 
and solving 
them one can obtain expressions for $S$ and $U$. With the known $U$,  
equation~(\ref{met11}) becomes the determining equation for the function $R$.  
Solving the latter one can get an explicit form for $R$.  
Compatible solutions to equations~(\ref{met9})-(\ref{met11}) can also be obtained 
in alternate ways, the details of which are given in \S3.

Now the functions $R,U$ and $S$ have to satisfy an extra set of constraints, 
that is, equations~(\ref{met12})-(\ref{met14}). Suppose a compatible solution 
satisfying all the equations have 
been found then the functions $R,\;U$ and $S$ fix the differential 
invariant $I(t,x,\dot{x},\ddot{x})$ by the relation 
\begin{align}
 I(t,x,\dot{x},\ddot{x}) & = r_1 - r_2 - \int \left\{RU+\frac{d}{d\dot{x}} \left[r_1 -
   r_2\right]\right\}d\dot{x}
  \nonumber\\
  &-\int\left\{R+\frac{d}{d\ddot{x}} \left[r_1 - r_2
   -\int \left\{RU+\frac{d}{d\dot{x}} \left[r_1-
  r_2\right]\right\}d\dot{x}\right]\right\}d\ddot{x}, 
  \label{met15}
\end{align}
where 
\begin{align} 
r_1  = \int R(\phi+S\dot{x}+U\ddot{x})dt,\;\;\;
r_2  =\int  \left( RS+\frac{d}{dx}\int r_1\right) dx. \nonumber
\end{align}
Equation~(\ref{met15}) can be derived straightforwardly by integrating the 
equations~(\ref{met8}). Here it is to be noted that for every independent 
set $(S,U,R)$, equation~(\ref{met15}) defines an integral.

\subsection{Exploring the complete form of R: Theory}
From the above discussion, it is clear that equation~(\ref{met1}) may be
considered as completely integrable once we obtain 
three independent sets of the solutions $(S_i,U_i,R_i),\;i=1,2,3,$ 
which provide three independent integrals of motion through the relation 
(\ref{met15}). Here we note that since we are solving
equations~(\ref{met9})-(\ref{met11}) first and then check the compatibility of 
this solution with equations~(\ref{met12})-(\ref{met14}), one often meets the 
situation that all the solutions which satisfy 
equations~(\ref{met9})-(\ref{met11}) need not satisfy the constraints 
(\ref{met12})-(\ref{met14}) 
since equations~(\ref{met9})-(\ref{met14}) constitute an overdetermined system 
for the unknowns $R$, $S$ and $U$. 
In fact for a class of problems one often gets one or two sets of $S,U,R$ 
which satisfy all the equations~(\ref{met9})-(\ref{met14})
and another set(s) $(S,U,R)$ which satisfies only the first three equations 
and not the other, namely, (\ref{met12})-(\ref{met14}). In this situation we
find an interesting fact that one can use the integral derived from the set(s)
which satisfies all the six equations (\ref{met9})-(\ref{met14}) 
and deduce the other compatible solution(s) $(S,U,\hat{R})$ (definition of
$\hat{R}$ follows). For example, let the set $(S_3,U_3,R_3)$ be a 
solution of the determining equations~(\ref{met9})-(\ref{met11}) and not of the 
constraints~(\ref{met12})-(\ref{met14}).  After analysing several examples we 
find that one can make 
the set $(S_3,U_3,R_3)$ compatible by modifying the form of $R_3$ as
\begin{eqnarray}
\hat{R_3} & = F(t,x,\dot{x},\ddot{x})R_3,
\label{met101}
\end{eqnarray} 
where $\hat{R_3}$ satisfies equation~(\ref{met11}), so that we have
\begin{eqnarray} 
(F_t+\dot{x}F_x+\ddot{x}F_{\dot{x}}+\phi F_{\ddot{x}}) R_3+FD[R_3]
   =-FR_3(U_3+\phi_{\dot{x}}).
\label{met102}
\end{eqnarray}
Further, if $F$ is a constant of motion (or a function of it), then the first 
term on the left hand side
vanishes and one gets the same equation~(\ref{met11}) for $R_3$ provided $F$ is 
non-zero. That is, whenever $F$ is a constant of motion or a function of it then
the solution to (\ref{met11}) may provide only a factor of the complete solution
$\hat{R_3}$ without the factor $F$ in equations~(\ref{met101}). This general form
of $\hat{R}_3$ along with $S_3$ and $U_3$ can now provide complete solution to
equations~(\ref{met9}) - (\ref{met14}) as discussed below.
\subsection{Exploring the complete form of R: Method}
Now if the sets $(S_i,U_i,R_i),\;i=1,2,$ are found to satisfy the equations~(\ref{met9}) - 
(\ref{met14}) and the third set $(S_3,U_3,R_3)$ does not satisfy 
equations (\ref{met12}) - (\ref{met14}) then $F$ may be
a function of the integrals $I_i,\;i=1,2,$  derived from the sets 
$(S_i,U_i,R_i),\;i=1,2$. 
We need to find the explicit form of $F(I_1,I_2)$ in order to obtain the
compatible solution $(S_3,U_3,\hat{R_3})$. To do so let us find the 
derivatives of $\hat{R_3}$ with respect to $x$, $\dot{x}$ and $\ddot{x}$: 
\begin{align} 
&\hat{R_3}_x=(F_1'I_{1x}+F_2' I_{2x})R_3+FR_{3x},\;\;
\hat{R}_{3\dot{x}}=(F_1'I_{1\dot{x}}+F_2' I_{2\dot{x}})R_3
+FR_{3\dot{x}},\nonumber\\
&\hat{R}_{3\ddot{x}}=(F_1'I_{1\ddot{x}}+F_2' I_{2\ddot{x}})R_3
+FR_{3\ddot{x}}, 
\label{fint01}
\end{align}
where $F_1'=\frac{\partial F}{\partial I_1}$ and 
$F_2'=\frac{\partial F}{\partial I_2}$. Substituting equation~(\ref{fint01})
into equations~(\ref{met12})-(\ref{met14}), we have 		
\begin{subequations}
\begin{eqnarray}
\frac{(f_1F_1'+f_2F_2')}{f_3}=\frac{F}{R_3},\;\; 
\frac{(f_4F_1'+f_5F_2')}{f_6}=\frac{F}{R_3}, \;\;
\frac{ (f_7F_1'+f_8F_2')}{f_9}=\frac{F}{R_3},
\label{fint02}								
\end{eqnarray}
where
\begin{eqnarray} 
&f_1=(I_{1x}-I_{1\ddot{x}}S_3),\quad 
f_2=(I_{2x}-I_{2\ddot{x}}S_3),\quad 
f_3=(S_3R_{3\ddot{x}}+R_3S_{3\ddot{x}}-R_{3x})\nonumber\\
&f_4=(I_{1\dot{x}}-I_{1\ddot{x}}U_3),\quad
f_5=(I_{2\dot{x}}-I_{2\ddot{x}}U_3),\quad
f_6=(U_3R_{3\ddot{x}}+R_3U_{3\ddot{x}}-R_{3\dot{x}})\nonumber\\
&f_7=(S_3I_{1\dot{x}}-I_{1x}U_3),\;
f_8=(S_3I_{2\dot{x}}-I_{2x}U_3),
\qquad\qquad\qquad\qquad\qquad\qquad\quad\nonumber\\
&f_9=(R_3U_{3x}+U_3R_{3x}-R_3S_{3\dot{x}}-S_3R_{3\dot{x}}).
\qquad\qquad\qquad\qquad\qquad\qquad\qquad\;\;
\label{fint05}								
\end{eqnarray}
\end{subequations}  

Equation~(\ref{fint02}) represents an overdetermined 
system of equations for the unknown $F$. A simple way to solve this equation is to uncouple
it for $F_1'$ ($=\frac{\partial F}{\partial I_1}$) and 
$F_2'$ ($=\frac{\partial F}{\partial I_2}$) and solve the resultant
equations. For example, eliminating $F_2'$ from equation~(\ref{fint02}) we 
obtain equations for $F_1'$ in the form
\begin{align} 
\frac{R_3F_1'}{F}=\frac{(f_3f_5-f_2f_6)}{(f_1f_5-f_2f_4)}
=\frac{(f_3f_8-f_2f_9)}{(f_1f_8-f_2f_7)}
=\frac{(f_6f_8-f_5f_9)}{(f_4f_8-f_5f_7)}.
\label{fint03}								
\end{align}
On the other hand eliminating $F_1'$ from equation~(\ref{fint02}) we arrive 
equations for $F_2'$ in the form
\begin{align} 
\frac{R_3F_2'}{F}=\frac{(f_3f_4-f_1f_6)}{(f_2f_4-f_1f_5)}
=\frac{(f_3f_7-f_1f_9)}{(f_2f_7-f_1f_8)}
=\frac{(f_6f_7-f_4f_9)}{(f_5f_7-f_4f_8)}.
\label{fint04}								
\end{align}

It can be easily cheeked that the compatibility of the right three expressions
in equations~(\ref{fint03}) or (\ref{fint04}) give rise to relations which are
effectively nothing but the constraint equations~(\ref{met12})-(\ref{met14}) 
and so no new
constraint is added now. Consequently, equations (\ref{fint03}) and
(\ref{fint04}) can be written as
\begin{align} 
\frac{\partial F}{\partial I_1}=g(I_1,I_2)F\quad \mbox{and}\quad
\frac{\partial F}{\partial I_2}=h(I_1,I_2)F,
\label{fint04a}								
\end{align}
respectively, where $g(I_1,I_2)=\frac{1}{R_3}\frac{(f_3f_5-f_2f_6)}
{(f_1f_6-f_3f_4)}$
and $h(I_1,I_2)=\frac{1}{R_3}\frac{(f_3f_4-f_1f_6)}{(f_2f_4-f_1f_5)}$. 
Now we can solve
equations (\ref{fint04a}) and obtain the form of $F(I_1,I_2)$. This is
demonstrated for several examples in the following sections explicitly.
Once $F$ is known we can obtain the complete solution
$\hat{R_3}$ from which, along with $S_3$ and $U_3$, the third 
integral $I_3$ can be constructed. {\it Thus with the explicit forms of the 
three integrals of motion, the complete integrability of equation~(\ref{met1}) 
is guaranteed}.

Finally, if the set $(S_1,U_1,R_1)$ alone is found to satisfy the 
equations~(\ref{met9}) - (\ref{met14}) and the second set $(S_2,U_2,R_2)$ 
also does not satisfy 
equations (\ref{met12}) - (\ref{met14}), then $F$ may be
a function of the integral $I_1$ alone which was derived from the set 
$(S_1,U_1,R_1)$. 
We need to find the explicit form of $F(I_1)$ in order to obtain the
compatible solution $(S_2,U_2,\hat{R_2})$. To do so let us find the 
derivatives of $\hat{R_2}$ with respect to $x$, $\dot{x}$ and $\ddot{x}$: 
\begin{align} 
\hat{R_2}_x=F_1'I_{1x}R_2+FR_{2x},\;\;
\hat{R}_{2\dot{x}}=F_1'I_{1\dot{x}}R_2
+FR_{2\dot{x}},\;\;
\hat{R}_{2\ddot{x}}=F_1'I_{1\ddot{x}}R_2+FR_{2\ddot{x}}, 
\label{fint01a}
\end{align}
where $F_1'=\frac{\partial F}{\partial I_1}$. Substituting 
equation~(\ref{fint01a})
into equations~(\ref{met12})-(\ref{met14}), we have 		
\begin{subequations}
\begin{eqnarray}
\frac{R_2F_1'}{F}=\frac{s_2}{s_1} 
=\frac{s_4}{s_3} 
=\frac{s_6}{s_5},
\label{fint06}								
\end{eqnarray}
where
\begin{eqnarray} 
s_1&=(I_{1x}-I_{1\ddot{x}}S_2),\quad \quad 
s_2=(S_2R_{2\ddot{x}}+R_2S_{2\ddot{x}}-R_{2x}),
\qquad \qquad \quad \nonumber\\
s_3&=(I_{1\dot{x}}-I_{1\ddot{x}}U_2),\quad\quad 
s_4=(U_2R_{2\ddot{x}}+R_2U_{2\ddot{x}}-R_{2\dot{x}}),
\qquad \qquad\quad \nonumber\\
s_5&=(S_2I_{1\dot{x}}-I_{1x}U_2),\quad
s_6=(R_2U_{2x}+U_2R_{2x}-R_2S_{2\dot{x}}-S_2R_{2\dot{x}}).
\label{fint06a}								
\end{eqnarray}
\end{subequations}
One can again check that the compatibility of the right three expressions in 
equation (\ref{fint06}) leads to a condition, which is deducible from
(\ref{met12})-(\ref{met14}), and so no new condition is introduced in reality.
Then rewriting equation (\ref{fint06}) as
\begin{align} 
\frac{\partial F}{\partial I_1}=\bigg(\frac{1}{R_2}\frac{s_2}{s_1}\bigg)F
=g(I_1)F,
\label{fint06b}								
\end{align}
and solving it, one can obtain an explicit expression for $F$. 
Once $F$ is known we can construct the complete form of $\hat{R_2}$ from
which, along with $S_2$ and $U_2$, the second integral of motion can be
obtained. Once $I_1$ and $I_2$ are known, we can proceed to find the third
compatible set $(S_3,U_3,\hat{R_3})$ as before and obtain
the third integral $I_3$
to establish complete integrability.

\section{Methods of finding the explicit form of $R$}
In \S2 we outlined the method of solving the determining equations. However, in
practice, it is difficult to solve equations (\ref{met9})-(\ref{met11})
straightforwardly as they constitute a set of coupled first order nonlinear 
partial differential equations (PDEs). So one has to look
for some intuitive ideas to solve these equations. We solve them and
obtain the forms $R,U$ and $S$ in the following way. For this purpose, we
observe the important fact that when
we rewrite the coupled equations (\ref{met9})-(\ref{met11}) into an equation 
for a single variable, namely, $R$, the resultant equation turns out to be a 
linear PDE. Then we solve this "$R$ equation" with a suitable ansatz 
(say polynomial
or rational in $\ddot{x}$). Once '$R$' is found the remaining functions $U$ and
$S$ can be easily deduced.


As noted, rewriting equations~(\ref{met9})-(\ref{met11}) we arrive at a  
third order linear partial differential equation for $R$ in the form
\begin{eqnarray}
 D^3[R]-D^2[R\phi_{\ddot{x}}]+D[R\phi_{\dot{x}}]-\phi_x R=0,\quad
D=\frac{\partial}{\partial{t}}+\dot{x}\frac{\partial}{\partial{x}}
+\ddot{x}\frac{\partial}{\partial{\dot{x}}}
+\phi\frac{\partial}{\partial{\ddot{x}}}.\label{poly01}
\end{eqnarray}
Solving equation~(\ref{poly01}) 
with a suitable ansatz in $\ddot{x}$ is relatively easier in many cases than 
solving equations~(\ref{met9})-(\ref{met11}). Once the explicit form of $R$ is 
obtained, $U$ can be deduced from equation~(\ref{met11}) as 
\begin{eqnarray} 
U=-(\frac{D[R]}{R}+\phi_{\ddot{x}}),
\label{poly02}
\end{eqnarray}
from which $S$ can be fixed by using equation~(\ref{met10}). Now if this set 
$(S,U,R)$ forms a compatible set for the remaining equations (\ref{met12})-
(\ref{met14}) then the corresponding integral $I$ can be found
using equation~(\ref{met15}). To illustrate this idea let us look into the
following examples.

\subsubsection{Example:1}   
Let us begin with a simple example, namely a linear third order ODE,
\begin{eqnarray}            
\dddot{x}+\lambda x=0,
\label{cat101}
\end{eqnarray}
where $\lambda$ is an arbitrary parameter.    
Substituting $\phi =-\lambda x$ into (\ref{poly01}), we get the following
equation for $R$ 
\begin{eqnarray}
D^3[R]+\lambda R=0,\;\;D=\frac{\partial}{\partial{t}}+
\dot{x}\frac{\partial}{\partial{x}}+\ddot{x}\frac{\partial}{\partial{\dot{x}}}
-\lambda x\frac{\partial}{\partial{\ddot{x}}}. \label{cat102}
\end{eqnarray}
We now assume an ansatz for $R$ in the form 
\begin{eqnarray} 
R = a(t,x,\dot{x})+b(t,x,\dot{x})\ddot{x},
\label{cat102a}
\end{eqnarray}
where $a$ and $b$ are arbitrary functions of $t,\;x$ and $\dot{x}$.
Substituting (\ref{cat102a}) into (\ref{cat102}) and equating the coefficients 
of different powers of $\ddot{x}$ to zero we get a set of linear partial 
differential 
equations for the variables $a$ and $b$.  Solving them one obtains three
nontrivial solutions,
\begin{eqnarray}
 R_1=-e^{kt}, \quad R_2=-(2\ddot{x}+k\dot{x}-k^2x)e^{-kt},\quad
R_3=-\frac{\sqrt{3}k}{2}(\dot{x}+kx)e^{-kt}, 
\label{cat103}
\end{eqnarray}
where $k^3=\lambda$. Now substituting the form of $R_i$'s $,i=1,2,3,$ 
separately into the equation (\ref{poly02}), we get
\begin{align}
U_1=-k,\quad U_2=\frac{(2k^2\dot{x}+k\ddot{x}+k^3x)}{(2\ddot{x}+k\dot{x}-k^2x)},
\quad U_3=-\frac{(\ddot{x}-k^2x)}{(\dot{x}+kx)}. 
\label{cat103a}
\end{align}
Now substituting the forms of $U_i$'s, $i=1,2,3,$ into (\ref{met10}) one can fix
the forms of $S_i$'s, $i=1,2,3,$ as
\begin{align}
S_1=k^2,\quad
S_2=\frac{2k^4x+k^3\dot{x}-k^2\ddot{x}}{2\ddot{x}+k\dot{x}-k^2x},\quad 
S_3=-k\frac{(\ddot{x}+k\dot{x})}{(\dot{x}+kx)}. \label{cat103b}
\end{align}
As a consequence now we have three sets of independent solutions for the 
equation 
(\ref{met9})-(\ref{met11}). Now we check the compatibility of these solutions 
with the remaining equations (\ref{met12})-(\ref{met14}).

We find that the solutions $(S_1,U_1,R_1)$ and $(S_2,U_2,R_2)$ 
satisfy the equations~(\ref{met12}) - (\ref{met14}), and become
compatible solutions. Substituting the forms $(S_1,U_1,R_1)$ and $(S_2,U_2,R_2)$
separately into equation (\ref{met15}) and evaluating the integrals we get
\begin{align} 
I_1& =\frac{1}{3k^2}\bigg(\ddot{x}-k\dot{x}+k^2x\bigg)e^{kt}, \label{cat111}\\
I_2&=\frac{2}{3k^2}\bigg(\ddot{x}^2+k^2\dot{x}^2+k^4x^2+k\dot{x}\ddot{x}
-k^2x\ddot{x}+k^3x\dot{x}\bigg)e^{-kt}.\label{cat112}
\end{align}
However, the set $(S_3,U_3,R_3)$ does not satisfy the extra 
constraints (\ref{met12})-(\ref{met14}) which means that  
the form of $R_3$  may not be the 'complete form' 
but might be a factor of the complete form. As mentioned in \S2, in order 
to recover the complete form $\hat{R_3}$ one may assume that  
$\hat{R_3} = F(I_1,I_2)R_3$,
where $F(I_1,I_2)$ is a function of the integrals $I_1$ and $I_2$. Now we have
to determine the 
form of $F(I_1,I_2)$ explicitly and for this purpose we proceed as follows.  
Substituting 
\begin{eqnarray}
\hat{R_3}  = F(I_1,I_2)R_3 = 
-\bigg[\frac{\sqrt{3}k}{2}(\dot{x}+kx)e^{-kt}\bigg]F(I_1,I_2),
\label{scat109a}
\end{eqnarray}
into equations~(\ref{fint03}) and (\ref{fint04}), along with (\ref{fint05}), we 
obtain two equations for 
$F$ as,
\begin{equation}
F_1'=0,\quad
I_2F_2'+F = 0,
\label{new1}
\end{equation}
where $F_1'$ and $F_2'$ denote partial derivatives of $F$ with respect to 
$I_1$ and $I_2$, respectively. Upon integrating (\ref{new1}) we get, 
$F = \frac {1}{I_2}$, (the integration constants are set to zero for simplicity)
which fixes the form of $\hat{R_3}$ as  
\begin{eqnarray}         
\hat{R_3} = \frac{R_3}{I_2}=-\frac{\sqrt{3}k}{2}\frac{(\dot{x}+kx)}
{(\ddot{x}^2+k^2\dot{x}^2+k^4x^2+k\dot{x}\ddot{x}-k^2x\ddot{x}+k^3x\dot{x})}. 
\label{scat110}  
\end{eqnarray}
Now one can easily check that this set $(S_3,U_3,\hat{R_3})$ is a compatible 
solution for the equations~(\ref{met9})-(\ref{met14}) which in turn provides
$I_3$ through the relation (\ref{met15}) as
\begin{eqnarray} 
I_3=-\frac{\sqrt{3}k}{2}t+tan^{-1}\bigg[\frac{\ddot{x}-k\dot{x}-2k^2x}
{\sqrt{3}(\ddot{x}+k\dot{x})}\bigg].
\label{scat111}
\end{eqnarray}
Using the explicit forms of the integrals $I_1,\; I_2$ and $I_3$, the 
solution to equation~(\ref{cat101}) can be deduced directly as
\begin{eqnarray}
x(t)=I_1e^{-kt}+\sqrt{I_2}e^{\frac{k}{2}t}\cos\bigg(\frac{\sqrt{3}k}{2}t
+I_3\bigg).
\label{scat112}
\end{eqnarray}
The result exactly coincides with the solution presented in Polyanin 
\& Zaitsev (2000).

\subsubsection{Example:2} 
The applicability of this method to nonlinear ODEs can be illustrated by
considering an equation of the form 
\begin{eqnarray} 
\dddot{x}=\frac{\ddot{x}^2}{\dot{x}}+\frac{\dot{x}\ddot{x}}{x}. \label {tex11}
\end{eqnarray}
Equation (\ref{tex11}) is a sub-case of the general form of a scalar third order
ODE which is invariant under the generators of time translation and rescaling 
(Feix \textit{et al.} 1997; Polyanin \& Zaitsev 2000). A sub-case of equation 
(\ref{tex11}), namely,
$\dddot{x}-c\frac{\ddot{x}^2}{\dot{x}}=0$, has been considered by both 
Bocharov \textit{et al.} (1993) and Ibragimov \& Meleshko (2005) to show that 
it can be linearized to a linear third order
ODE through a contact transformation. On the other hand, Euler \& Euler (2004)
have considered the equation $\dddot{x}-\frac{\dot{x}\ddot{x}}{x}=0$ and 
showed that it can be linearized through Sundman transformation 
(see \S5$\,c\,$ below). 
Here we consider the combined form (\ref{tex11}) and derive integrating factors,
integrals of motion and the general solution of this equation. Further, we show
that the equation (\ref{tex11}) itself can be linearized by the generalized
linearizing transformation (GLT) (see \S5$\,d\,$ below).

As before, substituting 
$\phi =\frac{\ddot{x}^2}{\dot{x}}+\frac{\dot{x}\ddot{x}}{x}$ into 
(\ref{poly01}), we get the following linear partial differential equation for
$R$ as
\begin{eqnarray}
D^3[R]-D^2[(\frac{2\ddot{x}}{\dot{x}}+\frac{\dot{x}}{x})R]
-D[(\frac{\ddot{x}^2}{\dot{x}^2}-\frac{\ddot{x}}{x})R]
+\frac{\dot{x}\ddot{x}}{x^2} R=0,\label{tex12}
\end{eqnarray}
where $D$ is defined by equation (\ref{opp}). Now substituting the ansatz 
(\ref{cat102a}) into (\ref{tex12}) and proceeding as
before, we get 
\begin{eqnarray}
 R_1=-\frac{1}{\dot{x}x},\quad R_2=\frac{x}{\dot{x}},\quad
R_3=\frac{t\dot{x}^2-x(\dot{x}+t\ddot{x})}{2x\dot{x}^2}. 
\label{tex13}
\end{eqnarray}
Following the ideas given in Example 1, one can deduce the corresponding forms
of $S_i$'s and $U_i$'s, $i=1,2,3,$ as
\begin{align}
 S_1&=-\frac{\ddot{x}}{x},\quad  
 &U_1=-\frac{\ddot{x}}{\dot{x}},\qquad \qquad \qquad \qquad \\
S_2&=\frac{\ddot{x}}{x},\quad 
&U_2=\frac{-2\dot{x}}{x}-\frac{\ddot{x}}{\dot{x}},\qquad \qquad \quad\;\;\\ 
S_3&=\frac{\dot{x}\ddot{x}(x+t\dot{x})}{x(-t\dot{x}^2
+x(\dot{x}+t\ddot{x}))}, \quad 
&U_3=\frac{x\ddot{x}(2\dot{x}+t\ddot{x})}{\dot{x}(t\dot{x}^2
-x(\dot{x}+t\ddot{x}))}.\; \;\;\;\;
\label{tex14}
\end{align}
The solutions $(S_i,U_i,R_i),\;i=1,2,$ satisfy the constraints 
(\ref{met12})-(\ref{met14}) so that they lead to first and second integrals
of the form
\begin{align}
I_1=\frac{\ddot{x}}{\dot{x}x}, \quad
I_2= \frac{2\dot{x}^2-x\ddot{x}}{\dot{x}}.\label{tex16}
\end{align}
In the present case also
the set $(S_3,U_3,R_3)$ does not satify the extra constraints and so one has
to explore the complete form of $\hat{R_3}$. To do so we proceed
as before and obtain the forms of $F$ and $\hat{R_3}$ as
$F = \frac {1}{\sqrt{I_1I_2}}$ and  
\begin{eqnarray} 
\hat{R_3} = \frac{t\dot{x}^2-x(\dot{x}+t\ddot{x})}
{2(\sqrt{I_1I_2})x\dot{x}^2}, 
\label{tex20}  
\end{eqnarray}
where the explicit forms of $I_1$ and $I_2$ are given in equation (\ref{tex16}).
Now one can check that the set $(S_3,U_3,\hat{R_3})$ satisfies all the six 
equations (\ref{met9})-(\ref{met14}) and furnishes the third integral $I_3$ 
through the relation (\ref{met15}) as 
\begin{eqnarray} 
I_3=-\frac{1}{2}(\sqrt{I_1I_2})t+tan^{-1}\sqrt{\frac{I_1}{I_2}}x.
\label{tex21}
\end{eqnarray}
Using the explicit forms of the integrals $I_1,\; I_2$ and $I_3$, the 
solution to equation~(\ref{tex11}) can be deduced directly as
\begin{eqnarray}
x(t)=\sqrt{\frac{I_2}{I_1}}\tan \bigg[\frac{1}{2}(\sqrt{I_1I_2}t+2I_3)\bigg].
\label{tex22}
\end{eqnarray}
As can be seen from equation~(\ref{tex20}) the complete compatible solution 
$\hat{R_3}$ has $\ddot{x}$ term which appear inside the square root sign. This
form of $\hat{R_3}$ can be explored only by making a suitable ansatz. Moreover
one may also face more difficulties in solving the determining equations~
(\ref{met9})-(\ref{met11}). In such complicated situations 
the complete solution $\hat{R}$ can be obtained by using our procedure.

\subsubsection{Example:3}
Let us consider a Chazy class of equation of the form 
(Halburd 1999; Mugan \& Jrad 2002; Euler \& Euler 2004)
\begin{eqnarray} 
\dddot{x}+4\alpha x\ddot{x}+3\alpha \dot{x}^2+6\alpha^2x^2\dot{x}
+\alpha^3x^4=0. \label {rex11}
\end{eqnarray}
Substituting $\phi =-(4kx\ddot{x}+3k\dot{x}^2+6k^2x^2\dot{x}+k^3x^4)$ into 
(\ref{poly01}), we get 
\begin{align}
D^3[R]+4\alpha D^2[xR]-6\alpha D[(\dot{x}+\alpha x^2)R]
+4\alpha (\ddot{x}+\alpha^2x^3+3\alpha x\dot{x})R=0.\label{rex12}
\end{align}
To solve the above equation we assume the same ansatz (\ref{cat102a}) for
the variable $R$. However, substituting this ansatz into equation~(\ref{rex12})
and solving the resultant equation we obtain only trivial solution. So we seek 
a rational form of ansatz for $R$ in the form
\begin{align}
R=\frac{a(t,x,\dot{x})+b(t,x,\dot{x})\ddot{x}}
{(c(t,x,\dot{x})+d(t,x,\dot{x})\ddot{x})^r} \label{rex14b},
\end{align}
where $r$ is an arbitrary number. Substuting the equation~(\ref{rex14b}) into 
equation~(\ref{rex12}) and solving the resultant partial differential equations 
we get
\begin{align}
 &R_1=\frac{\dot{x}+\alpha x^2}{(\alpha^2x^3+3\alpha x\dot{x}+\ddot{x})^2},\;\;
 R_2=\frac{t(-2x+\alpha tx^2+t\dot{x})}
 {(\alpha^2x^3+3\alpha x\dot{x}+\ddot{x})^2},\nonumber\\
 &R_3=\frac{t(3-3\alpha tx+\alpha^2t^2x^2+\alpha t^2\dot{x})}{(\alpha^2x^3
+3\alpha x\dot{x}+\ddot{x})^2}. 
\label{rex14e}
\end{align}
The corresponding forms of $S_i$'s and $U_i$'s $i=1,2,3,$ are
 \begin{align}
 &U_1=\frac{2\alpha^2x^3-\ddot{x}}{\alpha x^2+\dot{x}},
 \qquad\;S_1=\frac{\alpha(\alpha^2x^4+3\dot{x}^2-2x\ddot{x})}{\alpha x^2+\dot{x}},
 \nonumber\\
 &U_2=\frac{2x-6\alpha tx^2+2\alpha^2t^2x^3-t^2\ddot{x}}
 {t(-2x+\alpha tx^2+t\dot{x})},\label{rex13c}\\
 &S_2=\frac{2\alpha x^2-4\alpha^2tx^3+\alpha^3t^2x^4-2\dot{x}
  +3\alpha t^2\dot{x}^2+2t\ddot{x}(1-\alpha tx)}{t(-2x+\alpha tx^2+t\dot{x})},
  \nonumber\\
 &U_3=-\frac{3-12\alpha tx+9\alpha^2t^2x^2-2\alpha^3t^3x^3+\alpha t^3\ddot{x}}
{ t(3-3\alpha tx+\alpha^2t^2x^2+\alpha t^2\dot{x})},\nonumber\\
&S_3=\frac{(\alpha(12\alpha tx^2-6\alpha^2t^2x^3+\alpha^3t^3x^4+3t(2\dot{x}
 +\alpha t^2\dot{x}^2+t\ddot{x})-2x(3+\alpha t^3\ddot{x})))}{t(3-3\alpha tx
 +\alpha^2t^2x^2+\alpha t^2\dot{x})}.\nonumber 
\end{align}

Now we proeced to find the integrals of motion. First we note that the functions
$(S_1,U_1,R_1)$ satisfy the 
constraints~(\ref{met12})-(\ref{met14}) and hence they are compatible. 
Thus substituting them into (\ref{met15}) the first integral $I_1$ is fixed easily 
as 
\begin{align}
I_1&=-t+\frac{\alpha x^2+\dot{x}}
{\alpha^2x^3+3\alpha x\dot{x}+\ddot{x}}.\label{rex15}
\end{align}
However, the set $(S_2,U_2,R_2)$ (and so also $(S_3,U_3,R_3)$) does not 
satisfy the extra 
constraints (\ref{met12})-(\ref{met14}) which means   
the form of $R_2$  may not be the 'complete form' 
but might be a factor of the complete form. As mentioned in \S2, in order 
to recover the complete form $\hat{R_2}$ one may assume that  
$\hat{R_2}  = F(I_1)R_2$.
Here $F(I_1)$ is a function of the integral $I_1$. Now we have
to determine the 
form of $F(I_1)$ explicitly and for this purpose we proceed as follows.  
Substituting 
\begin{eqnarray}
\hat{R_2} & = F(I_1)R_2 = 
\frac{t(-2x+\alpha tx^2+t\dot{x})}{(\alpha^2x^3+3\alpha x\dot{x}+\ddot{x})^2}
F(I_1),
\label{rex15b}
\end{eqnarray}
into equation~(\ref{fint06}), we obtain $I_1F_1'+2F = 0$,
where $F_1'$ denotes differentiation with respect to 
$I_1$. Upon integrating the latter we get $F = I_1^{-2}$,
 (the integration constant is set to zero)
which fixes the form of $\hat{R_2}$ as  
\begin{eqnarray}         
\hat{R_2} = \frac{t(-2x+\alpha tx^2+t\dot{x})}
{(\alpha x^2-\alpha^2tx^3+\dot{x}-3\alpha tx\dot{x}-t\ddot{x})^2}. 
\label{rex15e}  
\end{eqnarray}
Now one can easily check that this set $(S_2,U_2,\hat{R_2})$ is a compatible 
solution for the equations~(\ref{met9})-(\ref{met14}) which in turn provides
$I_2$ through the relation (\ref{met15}) as
\begin{eqnarray} 
I_2=-\frac{-2\alpha tx^2+\alpha^2t^2x^3+x(2+3\alpha t^2\dot{x})+t(-2\dot{x}
+t\ddot{x})}{(\alpha x^2-\alpha^2tx^3+\dot{x}
-3\alpha tx\dot{x}-t\ddot{x})}.\label{rex16}
\end{eqnarray}

As in the previous examples, the set $(S_3,U_3,R_3)$ does not satisfy the 
constraints (\ref{met12})-(\ref{met14}) and hence one should seek a complete
form for $R_3$, which we denote as $\hat{R_3}$, in the form
\begin{eqnarray}
\hat{R_3} & = F(I_1,I_2)R_3 = 
F(I_1,I_2)\frac{t(3-3\alpha tx+\alpha^2t^2x^2+\alpha t^2\dot{x})}
{(\alpha^2x^3+3\alpha x\dot{x}+\ddot{x})^2}.
\label{rex17}
\end{eqnarray}
Substituting (\ref{rex17}) into equations~(\ref{fint03}) and (\ref{fint04}), 
we obtain the following equations for $F$, that is, 
$I_1F_1'+2F=0,\;\;F_2'= 0$,
where again $F_1'=\frac{\partial F}{\partial I_1}$ and $F_2'=\frac{\partial
F}{\partial I_2}$. 
Upon integrating the equations we get the explicit form of $F$ as 
$F = \frac {1}{I_1^2}$,
which in turn fixes the form of $\hat{R_3}$ as  
\begin{eqnarray}         
\hat{R_3} = \frac{t(3-3\alpha tx+k^2t^2x^2+\alpha t^2\dot{x})}
{3\alpha(\alpha x^2-\alpha^2tx^3+\dot{x}-3\alpha tx\dot{x}-t\ddot{x})^2}. 
\label{rex20}  
\end{eqnarray}
Now the set $(S_3,U_3,\hat{R_3})$ satisfies all the six equations 
(\ref{met9})-(\ref{met14}) and the relation (\ref{met15}) gives the form of
third integral $I_3$ as 
\begin{eqnarray} 
I_3=\frac{(6+3\alpha^2t^2x^2-\alpha^3t^3x^3+3\alpha t^2\dot{x}-3\alpha tx(2
+\alpha t^2\dot{x})-\alpha t^3\ddot{x})}{6\alpha(\alpha x^2-\alpha^2tx^3+\dot{x}
-3\alpha tx\dot{x}-t\ddot{x})}.
\label{rex21}
\end{eqnarray}
Thus we have obtained the explicit forms of the integrals $I_1,\; I_2$ and 
$I_3$ and hence the 
solution to equation~(\ref{cat101}) is obtained directly as
\begin{eqnarray}
x(t)=\frac{\frac{\alpha t^2}{2}+I_1t+I_1I_2}{\frac{\alpha^2t^3}{6}
+\alpha I_1\frac{t^2}{2}+\alpha I_1I_2t+I_1I_3}.
\label{rex22}
\end{eqnarray}

Interestingly one can derive the solution (\ref{rex22}) through an alternate way
also. For example, instead of solving the '$R$ equation' with rational
ansatz,
one can look for equations in other variables, that is, either in $U$ or in $S$. For 
example, from equation (\ref{met10}) we get
\begin{eqnarray}
 S=-(D[U]+\phi_{\dot{x}}-U\phi_{\ddot{x}}-U^2).\label{rat01}
\end{eqnarray}
Substituting equation~(\ref{rat01}) into the equation~({\ref{met9}) we get a
nonlinear PDE for $U$:
\begin{eqnarray}
D^2[U]-3UD[U]-UD[\phi_{\ddot{x}}]+D[\phi_{\dot{x}}]
-\phi_{x}-\phi_{\dot{x}}\phi_{\ddot{x}}+\phi_{\ddot{x}}^2U \qquad \qquad
\nonumber\\ \qquad\qquad \qquad \qquad 
+2\phi_{\ddot{x}}U^2-\phi_{\dot{x}}U+U^3=0.\label{rat02}
\end{eqnarray}
Now one can look for solutions of equation (\ref{rat02}) with polynomial in 
$\ddot{x}$. Once $U$ is known one can make use of 
equations~(\ref{rat01}) and (\ref{met11}) to get the forms of corresponding $S$  
and $R$ respectively. It turns out that for some cases, like the present example
(\ref{rex11}) (see the actual forms of $U$ in equation (\ref{rex13c})), solving 
equation~(\ref{rat02}) is  
easier than solving equation~(\ref{poly01}) to solve. However, this can be
decided only by actual calculation.  

\section{Linearization}
In \S2 and \S3 we discussed the complete integrability of third order ODEs by
investigating sufficient number of integrals of motion. However, one can also
establish the complete integrability of the given nonlinear ODE by transforming
it into a linear free particle second order ODE or into a third order linear
ODE of the form $\frac{d^3w}{dz^3}=w'''=0$. 
Unlike the second order ODEs, the third order nonlinear ODEs can be linearized
through a wide class of transformations, namely, invertible point
transformation (Steeb 1993; Ibragimov \& Meleshko 2005), 
contact transformation 
(Duarte \textit{et al.} 1994; Bocharov \textit{et al.} 1993; 
Ibragimov \& Meleshko 2005), 
generalised Sundman transformation 
(Berkovich \& Orlova 2000; Euler \textit{et al.} 2003; Euler \& Euler 2004) and
their generalizations. In the 
following, we describe a procedure to deduce these transformations
from the first integral itself and illustrate our ideas with relevant examples.
\subsection{Transformation from third order nonlinear ODEs to second order 
free particle equation}
Let us suppose that the ODE (\ref{met1}) admits a first integral
\begin{eqnarray}  
I_1  =F(t,x,\dot{x},\ddot{x}), \label{the01}
\end{eqnarray}
where $F$ is a function of $t,x,\dot{x}$ and $\ddot{x}$ only. Now extending our
earlier proposal for second order ODEs (Chandrasekar \textit{et al.} 2005) to
the third order equations (\ref{met1}), let us split the 
function $F$ into a product of two functions such that one
involves a perfect differentiable function $\frac{d}{dt}G_1(t,x,\dot{x})$ and
another function $G_2(t,x,\dot{x},\ddot{x})$, that is,
\begin{eqnarray} 
I_1=F\left(\frac{1}{G_2(t,x,\dot{x},\ddot{x})}\frac{d}{dt}G_1(t,x,\dot{x})
\right).\label{the02}
\end{eqnarray}
Suppose $G_2$ is a total time derivative of another function, say,
$z$, that is $dz/dt=G_2(t,x,\dot{x},\ddot{x})$, then (\ref{the02})
can be further rewritten in the form 
\begin{eqnarray} 
I_1=F\left(\frac{1}{\frac{dz}{dt}}\frac{dG_1}{dt}\right)
=F\bigg(\frac{dG_1}{dz}\bigg).
\label{the02b}
\end{eqnarray}
Now identifying the function $G_1(t,x,\dot{x})=w$ as the new dependent variable,
equation (\ref{the02b}) can be recast in the form
$I_1=F(\frac {dw}{dz})$.
In other words, we obtain
\begin{eqnarray} 
\hat{I}_1=\frac {dw}{dz},
\label{the02d}
\end{eqnarray}
where $\hat{I}_1$ is a constant, from which one can get $\frac{d^2w}{dz^2}=0$. 
Rewriting $w$ and $z$ in terms of old variables, namely,
\begin{eqnarray} 
w = G_1(t,x,\dot{x}),\quad z = \int_o^t G_2(t',x,\dot{x},\ddot{x}) dt', 
\label{the03}
\end{eqnarray}
we can get a linearizing transformation to transform the third order nonlinear ODE
into the second order free particle equation.  Then to deduce the general
solution one has to carry out one more integration.
\subsection{Transformation from third order nonlinear equation (\ref{met1}) 
to third order linear ODE $w'''=0$}
Next, we aim to transform equation (\ref{met1}) into a third
order linear ODE and so we try to rewrite the first integral as a 
perfect second order derivative.
We note that this can be done when one is able to evaluate the 
following integral explicitly, 
\begin{eqnarray} 
\hat{w} = \int_o^t G_1(t',x,\dot{x})G_2(t',x,\dot{x},\ddot{x}) dt'
=G_3(t,x,\dot{x}),
\label{the03a}
\end{eqnarray}
where $G_1$ and $G_2$ are defined as in equation (\ref{the02}). Note that in the
function $G_3$, the $\ddot{x}$ dependence has been integrated out. The reason
for making such a specific decomposition is that in this case equation 
(\ref{the02})
can be rewritten as a simple second order ODE for the variable $\hat{w}$ (see
equation (\ref{the003b}) below). We pursued a similar kind of approach in the 
integrable
force-free Duffing van der Pol oscillator equation (Chandrasekar \textit{et al.}
2004) which has now been generalized in the present case. Here one 
can rewrite (\ref{the01}) as a perfect second order derivative 
as follows.
Differentiating (\ref{the03a}) with respect to $t$, we obtain
$\frac {d\hat{w}}{dt}=G_1G_2$.
Rewriting the left hand side in the form
\begin{eqnarray} 
\frac {dz}{dt}\frac {d\hat{w}}{dz}=G_1G_2,
\label{the003}
\end{eqnarray}
and using the identities  already used in equation (\ref{the03}), namely 
$\frac {dz}{dt}=G_2$ and $G_1=w$, in equation (\ref{the003}) the latter becomes
\begin{eqnarray} 
\frac{d\hat{w}}{dz}=w.
\label{the003b}
\end{eqnarray}
Differentiating (\ref{the003b}) with repect to $z$ and using the
identity $\frac{dw}{dz}=\hat{I}_1$ (vide equation (\ref{the02d})), one gets
$\frac{d^2\hat{w}}{dz^2}=\hat{I}_1$.
In other words, we have
\begin{eqnarray} 
\frac{d^3\hat{w}}{dz^3}=0, \label{the05b}
\end{eqnarray}
so that $\hat{w}$ and $z$ are the required transformation variables.

\subsection{The nature of transformations}
In the previous two subsections, \S4$\,a\,$ and \S4$\,b\,$, we demonstrated 
how to
construct linearizing transformations from the first integral and how they
effectively change the given third order nonlinear ODE to either second 
or third order linear equation. Depending upon the explicit form of
the transformations we can call them as point, contact,
generalised Sundman or generalized linearizing transformations
(GLT). To 
demonstrate how 
different kinds of transformation arise let us consider the
transformation, $\hat{w} = G_3,\; z = \int_o^t G_2 dt'$ (vide 
equations (\ref{the03a}) and (\ref{the03})), which takes the given nonlinear ODE
into a linear equation. Now, in the above transformation, 
suppose $z$ is a perfect differentiable
function and $\hat{w}$ and $z$ do not contain the variable $\dot{x}$ and
$\ddot{x}$, then 
we call the resultant transformation, namely, $\hat{w}=f_1(x,t)$ and $z=f_2(x,t)$,  
as a point transformation. Further, if the transformation admits the variable 
$\dot{x}$ also explicitly then it
becomes the contact transformation. In this case, we have $w=f_1(t,x,\dot{x})$
and $z=f_2(t,x,\dot{x})$. On the other hand, if 
$\hat{w}=G_3(t,x)$ and $z = \int_o^t G_2(t',x) dt'$ then the
transformation is said to be a generalised Sundman transformation. Note that 
in the latter case the new independent
variable is in an integral form. Besides the above, as pointed out earlier, 
we find that
there exists another kind of transformation, namely, generalized linearizing 
transformations (GLT), in which the new dependent and independent variables
take the form $\hat{w} = G_3(t,x,\dot{x})$ and $z = \int_o^t
G_2(t',x,\dot{x},\ddot{x}) dt'$, respectively. 

\subsection{Connection between the functions $G_1,\;G_2$ and $G_3$}
Finally we explore the connection between the functions $G_1,G_2$ and $G_3$. 
As we have seen above, for the given equation to be linearizable it should be
transformable to
the form (\ref{the003b}). Rewriting the latter in terms of the variables, $G_1,G_2$
and $G_3$ we get,
\begin{eqnarray} 
 G_{3t}+\dot{x}G_{3x}+\ddot{x}G_{3\dot{x}}
 =G_1(t,x,\dot{x})G_2(t,x,\dot{x},\ddot{x}).
\label{cont01}
\end{eqnarray}
Note that the left hand side of equation (\ref{cont01}) contains the variable
$\ddot{x}$ linearly. So the right hand side should
also be linear in $\ddot{x}$. Consequently, we can write
\begin{eqnarray} 
 G_2(t,x,\dot{x},\ddot{x})
 =G_{21}(t,x,\dot{x})\ddot{x}+G_{22}(t,x,\dot{x}).
\label{cont04}
\end{eqnarray}
Using (\ref{cont04}) we can rewrite equation (\ref{cont01}) in the form
\begin{eqnarray} 
\frac{\bigg(G_{3t}+\dot{x}G_{3x}\bigg)\bigg(1+\frac{\ddot{x}G_{3\dot{x}}}
 {G_{3t}+\dot{x}G_{3x}}\bigg)}{G_{22}\bigg(1
 +\frac{\ddot{x}G_{21}}{G_{22}}\bigg)} =G_1.
\label{cont03}
\end{eqnarray}
Since the right hand side is independent of $\ddot{x}$, we have from 
(\ref{cont03}) that
\begin{eqnarray} 
 G_{21}(G_{3t}+\dot{x}G_{3x})=G_{22}G_{3\dot{x}}\quad
\mbox{and} \quad G_1=\frac{G_{3\dot{x}}}{G_{21}},
\label{cont02}
\end{eqnarray}
It may be noted that a similar condition has been derived by Bocharov 
\textit{et al.} (1993) and Ibragimov \& Meleshko (2005) for the case $G_2$ is a
perfect differential function. In other words, our procedure indicates that more
generalized transformations are possible in the case of third order ODEs. By
imposing the condition (\ref{cont04}) it becomes clear that whatever be the type
of linearizing transformation, the new independent variable should be at the
maximum linear in $\ddot{x}$.

\subsection{Transformation to fourth and higher order linear ODEs} 
In the above sub-sections \S4$\,a\,$, \S4$\,b\,$, \S4$\,c\,$ and \S4$\,d\,$ we
can concentrated only on transforming a nonlinear third order ODE either to a
second order or third order linear equations only. However, one could also
linearize certain third order nonlinear ODEs to fourth order linear ODE. For
example the equation (\ref{rex11}) is linearizable to a fourth order ODE of 
the form $\frac{d^4w}{dz^4}=0$, under the nonlocal transformation  
$x=\frac{\dot{w}}{\alpha w}$. This is not an isolated example and one can 
linearize
a class equations through this procedure. Besides the above one can also consider
linearizing transformations in which the new dependent variable, $\hat{w}$, is a
non-local one. This choice leads us to classify another a large class of
equations which we leave it for future work. 
\section{Application}
In this section we consider specific examples to demonstrate the method given 
in \S4.
\subsection{Example 1: Point transformation}
Let us consider a nontrivial example which was discussed by 
Steeb (1993) in the context of invertible point transformations, namely,   
\begin{eqnarray} 
\dddot{x}+\frac{3\dot{x}\ddot{x}}{x}-3\ddot{x}-\frac{3\dot{x}^2}{x}+2\dot{x}=0. 
\label{fthe20}
\end{eqnarray}
The first integral, which can be obtained using the formulation in \S2, can be
written as
\begin{eqnarray} 
I_1 =\bigg(\dot{x}^2+x\ddot{x}-x\dot{x}\bigg)e^{-2t}. \label{the20}
\end{eqnarray}
Rewriting (\ref{the20}) in the form (\ref{the02}) we get
$I_1 =e^{-t}\frac {d}{dt}(x\dot{x}e^{-t})$, so that 
\begin{eqnarray}  
w=x\dot{x}e^{-t},\quad z=e^{t}. \label{the22}
\end{eqnarray}
As we noted earlier, one could transform (\ref{fthe20}) to the second order free
particle equation, $\frac{d^2w}{dz^2}=0$, by utilizing the transformation 
(\ref{the22}). Integrating the equation $\frac{d^2w}{dz^2}=0$ we get 
$w=I_1z+I_2$. Using (\ref{the22}) into this expression, the general solution 
of equation (\ref{fthe20}) can be obtained  (after an integration) as
\begin{eqnarray} 
x(t)=(I_1e^{2t}+I_2e^t+I_3)^{\frac{1}{2}},\label{the23c} 
\end{eqnarray}
where $I_i,\;i=1,2,3,$ are the integration constants.
 
Further, using the equation (\ref{the03a}) we get
$\hat{w} = \int x\dot{x}dt = \frac{x^2}{2}$.
Then we can directly check that the point transformation 
\begin{eqnarray} 
\hat{w}=\frac{x^2}{2}, \quad z=e^{t},\label{the23b}
\end{eqnarray}
transforms the equation (\ref{fthe20}) to the form  $\frac{d^3\hat{w}}{dz^3}=0$.
As we mentioned earlier, since $\hat{w}$ and $z$ involve only $x$ and $t$ 
they are just point transformations. Integrating the linear equation
$\frac{d^3\hat{w}}{dz^3}=0$ we get
\begin{eqnarray} 
\hat{w}=\frac{I_1}{2}z^2+I_2z+I_3. \label{tmet01}
\end{eqnarray}
Rewriting $\hat{w}$ and $z$ in equation (\ref{tmet01}) in terms of the original 
variables 
$x$ and $t$ by using the transformation (\ref{the23b}) we get the same solution as
equation (\ref{the23c}).

\subsection{Example 2: Contact transformation}
Let us consider an equation of the form 
\begin{eqnarray} 
\dddot{x}=\frac{x\ddot{x}^3}{\dot{x}^3}. 
\label{che20a}
\end{eqnarray}
Bocharov \textit{et al.} (1993) have shown that equation (\ref{che20a}) can be
linearized through contact transformation. However, the explicit linearizing
transformation is yet to be reported which is also a difficult problem. Here we
derive the explicit form of the linearizing transformation through our
procedure.

The first integral can be easily deduced using the results of \S2 as 
\begin{eqnarray} 
I_1 =\frac{\dot{x}^2-x\ddot{x}}{\dot{x}\ddot{x}}. \label{che20}
\end{eqnarray}
Rewriting (\ref{che20}) in the form (\ref{the02}) we get
$I_1 =\displaystyle{\frac{\dot{x}}{\ddot{x}}\frac {d}{dt}(\frac{x}{\dot{x}})}$
so that we have $w=\frac{x}{\dot{x}}$ and $z=\log{\dot{x}}$. The latter transforms
equation (\ref{che20a}) to the second order free particle equation
$\frac{d^2w}{dz^2}=0$, so that $w=I_1z+I_2$,
where $I_1$ and $I_2$ are integration constants. Rewriting $w$ and $z$ in terms
of the old variables we get $x=(I_1\log(\dot{x})+I_2)\dot{x}$. Unlike the
earlier example it is difficult to integrate this equation further and obtain 
the general solution. Therefore one can look for variables which
transform the third order nonlinear ODE (\ref{che20a}) to a third order linear 
ODE so that the 
non-trivial
integration can be avoided. Now using the equation (\ref{the03a}) we get 
$\hat{w} = \int \frac{x\ddot{x}}{\dot{x}^2}dt = t-\frac{x}{\dot{x}}$.
The new variables 
\begin{eqnarray} 
\hat{w}=\frac{t\dot{x}-x}{\dot{x}}, \quad z=\log{\dot{x}},
\end{eqnarray}
transform the equation (\ref{che20a}) to the 
form (\ref{the05b}). Unlike the earlier example, $\hat{w}$ and $z$ admit 
the
variable $\dot{x}$ explicitly and so they become contact transformation for the
given equation.

Integrating the linear third order equation (\ref{the05b}), we get 
$\hat{w}=\frac{I_1}{2}z^2+I_2z+I_3,$ 
where $I_i,\;i=1,2,3,$ are integration constants. Now replacing $\hat{w}$ and $z$
in terms of the old variables and using the previous result 
$x=(I_1\log(\dot{x})+I_2)\dot{x}$, one can obtain the general solution for the equation
(\ref{che20a}) in the form
\begin{eqnarray} 
x(t)=\bigg(-I_1\pm\sqrt{I_1^2+I_2^2-2I_1(I_3-t)}\bigg)
e^{-\frac{I_1+I_2\mp\sqrt{I_1^2+I_2^2-2I_1(I_3-t)}}{I_1}}. \label{cmet02}
\end{eqnarray}

\subsection{Example 3: Generalised Sundman transformation}
Next we consider the hydrodynamic type equation of the form 
(Berkovich \& Orlova 2000; Euler \& Euler 2004)
\begin{eqnarray} 
\dddot{x}=\frac{\ddot{x}\dot{x}}{x}, \label {sex21}
\end{eqnarray}
which admits a first integral in the form $I_1=\frac{\ddot{x}}{x}$ and the
latter can be rewritten as 
$I_1 =\frac{1}{x}\frac {d}{dt}\dot{x}=\frac{dw}{dz}$, from which we identify 
$w=\dot{x}$ and $dz=xdt$.
By utilizing the new variables, one can transform (\ref{sex21}) to the second order free
particle equation, $\frac{d^2w}{dz^2}=0$. 
However, from equation (\ref{the03a}) we get $\hat{w} = \int x\dot{x}dt = x^2$.
Then the Sundman transformation 
\begin{eqnarray} 
\hat{w}=x^2, \quad dz=x dt,\label {sex22}
\end{eqnarray}
transforms the equation (\ref{sex21}) to the 
form (\ref{the05b}), namely $\frac{d^3\hat{w}}{dz^3}=0$. 

To derive the solution we proceed as follows. Rewriting the first integral $I_1$
in the integral form we get, $I_1 =\frac{1}{x}\frac {d}{dt}\dot{x}\Rightarrow
\dot{x}=I_1\int xdt$. Now using the identity (\ref{sex22}) in the latter expression
we get $w=I_1z$.
From equation (\ref{the03a}) (for the present case $G_1=w$ and
$G_2=\frac{dz}{dt}$) we have
\begin{eqnarray} 
\hat{w}=\int w dz=\int I_1zdz=\frac{I_1}{2}z^2+I_2, \label {sex25}
\end{eqnarray}
where $I_2$ is the integration constant. Using (\ref{sex22}) in (\ref{sex25}) 
we obtain $x^2=\frac{I_1}{2}z^2+I_2$, 
which in turn
leads to a differential equation which connects the variables $z$ and $t$ in 
the form (using the relation $dz=xdt$)
\begin{eqnarray} 
dz=\sqrt{\frac{I_1}{2}z^2+I_2}dt. \label {sex27}
\end{eqnarray}
Integrating (\ref{sex27}), we obtain
$z= \sqrt{\frac{I_2}{2}}(e^{\sqrt{I_1}(t+I_3)}-e^{-\sqrt{I_1}(t+I_3)})$, where
$I_3$ is the integration constant. 
Substituting the latter in the relation
$x^2=\frac{I_1}{2}z^2+I_2$, we arrive at the general solution for 
(\ref{sex21}) in the form
\begin{eqnarray} 
x(t)=\sqrt{I_2} \cosh{\sqrt{I_1}(t+I_3)}.\label {sex28}
\end{eqnarray}
We note that the solution for the equation (\ref{sex21}) has been already 
derived in an alternate way from the Sundman transformation (\ref{sex22}) 
by Euler \& Euler (2004). However {\it the
procedure which we described in the above is new and can be used for more
general linearizing transformations also}, as we see below.

\subsection{Example 4: Generalized linearizing transformation (GLT)}
As we noted earlier \textit{some nonlinear ODEs can be linearized only through 
more general
nonlocal form of transformations}, which we designate here as generalized 
linearizing transformations (GLTs). We illustrate the GLT with the same example
discussed as Example 2 in \S3, which admits a first integral of the form
$I_1=\frac{\dot{x}x}{\ddot{x}}$ (vide equation (\ref{tex11})).
Rewriting this first integral in the form (\ref{the02}) we get
$I_1 =\frac{x}{\ddot{x}}\frac {d}{dt}(x),$
so that 
\begin{eqnarray}  
w=x,\quad dz=\frac{\ddot{x}}{x}dt,\label {gex11c}
\end{eqnarray}
which can be effectively used to transform the nonlinear ODE (\ref{tex11}) to the
equation $\frac{d^2w}{dz^2}=0$.
Using the equation (\ref{the03a}) we get 
$\hat{w} = \int \ddot{x}dt = \dot{x}$.
Then the GLT becomes 
\begin{eqnarray} 
\hat{w}=\dot{x}, \quad dz=\frac{\ddot{x}}{x}dt,\label {gex12}
\end{eqnarray}
which can be used to transform the equation (\ref{tex11}) to the form 
$\hat{w}'''=0$. Note that in the present case the new independent and
dependent variables admit $\ddot{x}$ and $\dot{x}$ terms, respectively and that
the transformation is nonlocal. {\it 
Indeed no
such linearizing transformations have been reported in the literature atleast to
our knowledge. We also now establish a method of finding the general solution for
this case}.

Integrating once the equation $\frac{d^2w}{dz^2}=0$ we get $w=I_1z$ from which
we obtain  
\begin{eqnarray} 
x=I_1z.\label {gex12a}
\end{eqnarray}
On the other hand equation (\ref{the03a}) provides us a relation (after using
(\ref{gex11c}) and (\ref{gex12}))
\begin{eqnarray}  
\dot{x}=\frac{I_1}{2}z^2+I_2.\label {gex12b}
\end{eqnarray}
Now using (\ref{gex12a}) in (\ref{gex12b}), we obtain 
\begin{eqnarray} 
\bigg(\frac{2I_1}{I_1z^2+2I_2}\bigg)dz=dt. \label {gex17}
\end{eqnarray}
The variables are now separated out and one can integrate
(\ref{gex17}) and obtain
\begin{eqnarray} 
z=\sqrt{\frac{2I_2}{I_1}}\tan\sqrt{\frac{I_2}{I_1}}(t+I_3).\label {gex18a}
\end{eqnarray}
Now substutiting (\ref{gex18a}) into (\ref{gex12a}) we get
\begin{eqnarray} 
x(t)=\sqrt{2I_1I_2}\tan\sqrt{\frac{I_2}{I_1}}(t+I_3).\label {gex18}
\end{eqnarray}
which is effectively the same as (\ref{tex22}).

Finally, we note that the procedure given above can be profitably utilized for 
other examples also which are linearized by GLTs.
\subsection{Example 5: An elementary nontrivial system of hydrodynamic type}
Finally, to show the importance of the GLT and how this transformation gives 
additional information about the
linearization of nonlinear third order ODEs we consider the following specific
example which was discussed in the literature (Berkovich 1996; 
Berkovich \& Orlova 2000), 
\begin{eqnarray} 
\dddot{x}=\frac{\ddot{x}\dot{x}}{x}-4\alpha x^2\dot{x},\;\;\alpha :\;parameter. 
\label {gex21}
\end{eqnarray}
Equation (\ref{gex21}) is nothing but the dynamical equation of the 
Euler-Poinsot case of a rigid body written in terms of a single variable 
(Berkovich 1996; Berkovich \& Orlova 2000). As we have seen earlier this equation is 
linearizable in the case $\alpha=0$ through generalized Sundmann
transformation. \textit{However, we wish to show here that the general equation 
(\ref{gex21}) itself is linearizable through the GLT}.

From the first integral $I_1=\frac{\ddot{x}}{x}+2\alpha x^2$ associated with equation
(\ref{gex21}),
one can identify the GLT
\begin{eqnarray} 
\hat{w}=x^2, \quad dz=\frac{2\dot{x}x}{\sqrt{\dot{x}^2+\alpha x^4}}dt,\label {gex22}
\end{eqnarray}
which transforms the equation (\ref{gex21}) to the form (\ref{the05b}). 
Note that for the choice $\alpha=0$ the independent variable becomes $dz=2xdt$ and so
it becomes the generalized Sundman transformation, equation (\ref{sex22}), 
identified in the literature 
(Berkovich \& Orlova 2000; Euler \& Euler 2004). Now following the 
steps given in Example 4 one can deduce the general solution for the equation
(\ref{gex21}) in terms of Jacobian elliptic function as
\begin{eqnarray} 
x(t)=\bigg(I_1(c-(c-b)sn^2[\sqrt{\alpha
I_1(c-a)}(t-t_0),m])+I_2\bigg)^{\frac{1}{2}},
\label {gex28}
\end{eqnarray}
where $a+b+c=\frac{(2I_1-3\alpha I_2)}{4\alpha I_1},\;
ab+ac+cb=\frac{(3\alpha I_2^2-2I_1I_2)}{4\alpha I_1^2},
\;abc=-\frac{I_2^3}{4I_1^3}$, $m^2=\frac{b-c}{a-c}$ and $I_2$ and $t_0$ are
integration constants.

\section{Conclusion}
In this paper we have discussed a method of finding the integrals of motion and 
general solution
associated with third order nonlinear ODEs through the modified PS method
by a nontrivial extension of our earlier work on second order ODEs 
(Chandrasekar \textit{et al.} 2005). We
illustrated the validity of the method with suitable examples. Further, we introduced a
technique which can be utilized to derive linearizing transformations from the
first integral. Interestingly, we showed that different types of
transformations, namely, point, contact, Sundman and generalized linearizing 
transformations can be derived in a unique way from the first integral. 
We also indicated a procedure to derive general solution for the third order 
ODEs when GLTs occur. We believe that the GLT introduced in this 
paper will be highly useful to tackle new systems such as equation 
(\ref{gex21}). Finally, the modified PS
method can also be extended to higher order ODEs and coupled systems of 
ODEs. As far as the linearization of higher order ODEs are concerned
it is still an open and challenging problem. As we pointed out in the
introduction one can unearth a wide variety of linearizing transformations for
the higher order ODEs besides formulating the necessary and sufficient condition
for linearizing these equations in each form of transformation. We hope to 
address some of these aspects shortly.	

The work of VKC is supported by Council of Scientific and Industrial Research 
in the form of a Senior Research
Fellowship.  The work of MS and ML forms part of a Department of Science 
and Technology, Government of India, sponsored research project.

\end{document}